\documentclass[runningheads]{llncs}

\usepackage[T1]{fontenc}
\usepackage{enumitem}
\usepackage{amsmath,amssymb,amsfonts}
\usepackage{graphicx}
\usepackage{textcomp}
\usepackage{url}

\usepackage{soul}
\usepackage{xcolor}
\sethlcolor{gray!25}

\usepackage{array}
\usepackage{booktabs}
\usepackage{multirow}

\graphicspath{{figures/}}

\begin{document}

\title{Adversarial Robustness in Smishing Detection: A Comparative
Analysis of Adversarial Fragility in Classical vs.\ Transformer-Based
Detection Systems}

\titlerunning{Adversarial Robustness in Smishing Detection}

\author{Denzel Chiuseni \and Athanase Bahizire \and Silva Hama \and Jema David Ndibwile}

\authorrunning{Chiuseni et al.}

\institute{Carnegie Mellon University Africa, Kigali, Rwanda \\
\email{\{dchiusen, abahizir, shama, jndibwil\}@andrew.cmu.edu}}

\maketitle

\begin{abstract}
Smishing detection systems are commonly trained and evaluated on clean, monolingual text. In low-resource settings, however, attackers frequently circumvent these systems through character obfuscation, cross-lingual code-switching, and structural perturbation. This study evaluates adversarial robustness for  five model architectures: three classical lexical models (Random Forest, XGBoost, CNN+BiLSTM) and two multilingual transformers (mBERT, XLM-RoBERTa), using a dataset of 27,037 messages. Classical models are subjected to black-box generic attacks, while transformers are evaluated with attention-guided targeting. Each model is tested across three attack types and intensity levels, with performance measured by the Robustness Degradation Ratio (RDR). The results reveal a distinct architectural boundary: classical models experience near-catastrophic failure under character obfuscation and structural perturbation (RDR up to 0.988), whereas transformers demonstrate significantly greater resilience (RDR up to 0.351), with structural perturbation representing their most pronounced vulnerability. Effect-size analysis (Cliff’s $\delta$) indicates a substantial difference between the two model categories. Within the transformer group, XLM-RoBERTa, despite achieving a higher clean-text baseline, exhibits greater degradation than mBERT. These findings demonstrate that clean-text performance is not a reliable predictor of adversarial robustness. Statistical validation using Mann-Whitney U and Friedman tests confirms that these patterns are attributable to model architecture rather than sampling. The results underscore the necessity for architecture-specific defences and frame smishing detection as an adversarial cybersecurity challenge rather than a static classification task.

\end{abstract}

\section{Introduction}
SMS continues to serve as a primary channel for transaction notifications, authentication, and service alerts within financial systems. Its low cost and extensive reach render it a frequent target for SMS-based phishing (smishing) attacks, which employ social engineering to acquire sensitive credentials such as PINs and OTPs. This threat is particularly pronounced in low-resource settings, where limited labelled data, linguistic diversity, and varying levels of user security awareness impede effective detection. Mobile money platforms are especially susceptible, as they rely on SMS for essential operations and serve large, linguistically diverse user bases. As detection systems advance, attackers have adopted increasingly sophisticated evasion tactics, including character substitution, code-switching, and structural manipulation, to circumvent traditional filters.
\section{Literature Review}

Current research on smishing detection emphasises algorithmic optimisation and classification reliability spanning from optimised machine learning ensembles ~\cite{mambina_classifying_2022,xu_malicious_2025} to advanced deep learning architectures ~\cite{mambina_uncovering_2024,haizam_analysing_2024}. These approaches are predicated on the assumption that SMS text stays linguistically clean, homogeneous, and monolingual. This reliance results in a strong reliance on lexical extraction for classifying malicious intent  ~\cite{mambina_classifying_2022,hosseinpour_complex-network_2024}. Even context-aware deep learning systems remain susceptible to token-level perturbations and distribution shifts, particularly within low-resource settings. Although these vulnerabilities are well documented in theory, existing systems have not been systematically evaluated under the adversarial conditions that characterise real-world SMS environments. Evasion tactics should therefore be regarded as persistent environmental variables rather than as anomalous edge cases.

Attackers exploit these vulnerabilities through multiple evasion strategies. \textit{Character-level obfuscation} introduces imperceptible noise, visually similar homoglyphs, non-printable characters, or intentional misspellings to alter the digital encoding of trigger words while maintaining human readability ~\cite{boucher_bad_2021,li_textbugger_2019}; Conventional systems are unable to distinguish benign typographical variation from targeted malicious noise, resulting in degraded detection capability and successful evasion of malicious payloads ~\cite{bajaj_deceiving_2023,shaikh_smishing_2025}. \textit{Structural and spacing perturbations}manipulate text boundaries by inserting or removing spaces, thereby bypassing the tokenisation phase and causing models to process payloads as out-of-vocabulary fragments ~\cite{morris_textattack_2020}. These anomalies integrate seamlessly into SMS traffic, where character limits already produce irregular spacing. \textit{Cross-lingual code-switching},based on the concept of the adversarial polyglot, fragments the semantic structure of malicious payloads across linguistic borders ~\cite{tan_code-mixing_nodate}. This technique conceals payloads within natural code-switched communication and can result in performance declines of up to 36\% in standard detection systems~\cite{asheshemi_adversarial_2025}, as typical algorithms are unable to reconstruct fractured intent across languages.

Collectively, the literature demonstrates a fundamental excessive dependence on the assumption that SMS threat environments are static and linguistically homogeneous. This perspective obscures an architectural dependence on surface-level lexical features, which the three identified attack strategies exploit while preserving malicious intent.

\section{Research Gap}
Smishing detection research has grown substantially in recent years ~\cite{mambina_classifying_2022,mambina_uncovering_2024,sankaine_english-swahili_2025}, but adversarial robustness is rarely treated as a primary evaluation criterion: models are trained and tested on clean, monolingual text, with evasion tactics discussed theoretically rather than tested empirically. Three specific gaps define the problem this study addresses.
\begin{itemize}
\item No prior work evaluates smishing detection models across multiple adversarial attack types at graded intensities within low-resource settings; 
\item No existing study compares adversarial fragility across the classical-to-transformer spectrum in this setting, leaving the boundary between TF-IDF feature collapse and transformer subword insulation empirically undrawn.
\item No prior work has tested whether clean-text performance reliably predicts adversarial robustness; if higher-performing models prove more fragile under targeted attack, optimising for accuracy offers a false sense of security.
\end{itemize}
This study fills all three gaps through a systematic evaluation using the RDR, applied across five model architectures, three attack types and intensity levels.

\section{Research Questions}
The central problem this study addresses is: \textit{ To what extent do contemporary smishing detection systems stay robust under adversarial evasion conditions, and does architectural design determine the nature and severity of that fragility?}
To address this, the following sub-questions are proposed:
\begin{itemize}
    \item \textbf{RQ1:} How does the performance of the smishing detection degrade under adversarial attack at varying intensities?

    \item \textbf{RQ2:} How do the architectural properties of classical and transformer-based detection systems determine differential robustness under adversarial evasion in a low-resource context?

    \item \textbf{RQ3:} Does clean-text classification performance serve as a reliable predictor of adversarial robustness, or can architecturally superior models exhibit greater fragility?
\end{itemize}

\subsection{Hypotheses}

The following hypotheses were tested to evaluate the proposed framework.

\subsubsection{Hypothesis 1: Cross-Tier Fragility}
\mbox{}\\
\textbf{H0:} There is no significant difference in RDR between classical lexical-based and transformer-based detection systems under high-intensity adversarial attack. 
\mbox{}\\
\textbf{Ha:} Classical lexical-based systems exhibit significantly higher RDR than transformer-based models under high-intensity conditions.

\subsubsection{Hypothesis 2: Performance-Robustness Decoupling}
\mbox{}\\
\textbf{H0:} Clean-text classification performance reliably predicts adversarial robustness.
\mbox{}\\
\textbf{Ha:} Clean-text performance does not predict adversarial robustness under attack conditions.

\subsubsection{Hypothesis 3: Perturbation Intensity and Non-Linear Degradation}
\mbox{}\\
\textbf{H0:} $F1_{degradation}$ scales linearly and predictably with increasing adversarial perturbation intensity. 
\mbox{}\\
\textbf{Ha:} Detection systems exhibit a non-linear robustness threshold, maintaining resilience against low-intensity noise but dropping sharply once intensity crosses a critical point.

\subsection{Contribution}
This study treats smishing detection as an adversarial security problem rather than a conventional classification task. It offers a systematic empirical comparison of how classical and transformer-based architectures degrade under adversarial conditions in a low-resource SMS context, quantifying architecture-specific failure points through principled robustness metrics under worst-case attack scenarios. Although the evaluation is conducted in an English-Swahili setting, the attack strategies are language-agnostic by design: homoglyph substitution operates across scripts, structural perturbation targets tokenisation boundaries common to any subword pipeline, and code-switching applies wherever natural language mixing occurs. The framework therefore transfers to other low-resource language pairs without methodological modification.

\section{Methodology}

This section outlines the systematic approach taken in this study. It describes how adversarial attacks were generated and applied to existing smishing detection systems.

\subsection{Dataset Construction and Preprocessing}
Two datasets were combined to construct the experimental corpus: the Kaggle Swahili SMS Dataset (1,508 Tanzanian Swahili messages) and an English-Swahili Dataset (62,647 messages spanning broader SMS traffic patterns). After merging and deduplication, the corpus contained 27,037 messages, of which 356 (1.3\%) were labelled malicious. Preprocessing covered lowercasing, whitespace normalisation, null filtering, and binary label standardisation (1 = malicious, 0 = benign). The dataset was split 80:20 using stratified sampling with random state 42, yielding approximately 71 malicious messages in the test partition.

This limited test size is a direct consequence of the low-resource environment under investigation. RDR mitigates its impact on the validity of the result by measuring relative degradation using F1 rather than aggregate accuracy, thereby isolating degradation regardless of sample size. However, the limited malicious test set is a constraint of this study and motivates future work on larger annotated corpora for resource-limited smishing evaluation.

\subsection{Candidate Models}
Five models were selected across two tiers to enable direct comparative evaluation of lexical versus semantic robustness under adversarial conditions.

\subsubsection{Classical Tier}
\begin{itemize}
    \item \textbf{Random Forest:} A TF-IDF $n$-gram ensemble classifier operating on bag-of-words features, chosen for its interpretable feature importances, which expose the trigger-token dependencies that adversarial attacks target.

    \item \textbf{XGBoost:} A gradient-boosted classifier on the same TF-IDF feature space, included for comparison across ensemble strategies within the lexical paradigm.

  \item \textbf{CNN+BiLSTM:} A hybrid architecture combining local $n$-gram extraction (CNN) with bidirectional context modelling (BiLSTM), more context-aware than bag-of-words models but still bound to whole-word tokens.
\end{itemize}

\subsubsection{Transformer Tier}
\begin{itemize}
    \item \textbf{mBERT:} A transformer trained on 104 languages using WordPiece subword tokenisation, fine-tuned on the merged corpus for binary classification. \cite{pires_how_2019}
    
    \item \textbf{XLM-RoBERTa:} A transformer using SentencePiece tokenisation with stronger low-resource coverage than mBERT, fine-tuned identically; its superior cross-lingual transfer suits code-switching detection \cite{conneau_unsupervised_2020}.
\end{itemize}

\subsection{Attack Strategy --- Classical Models}
Classical models (Random Forest, XGBoost, CNN+BiLSTM) are evaluated under a black-box threat model, in which the attacker has no access to internal parameters, gradients, or decision boundaries. These architectures classify on surface-level lexical representations and are consequently sensitive to perturbations at any token position, making generic black-box attacks well-suited to expose their brittleness. Adversarial examples are generated via the Gemini API with prompts designed to preserve human readability while degrading surface features, applied across three intensity levels: low, medium, and high.

Three attack strategies are employed. Character obfuscation substitutes ASCII characters with visually identical Unicode homoglyphs, zero-width non-printable characters, or deliberate misspellings, disrupting token matching while preserving readability. Structural perturbation inserts or removes spaces within and between tokens, or alters punctuation boundaries, forcing models to process payloads as out-of-vocabulary fragments; such irregularities blend readily into SMS traffic, where character constraints already produce non-standard spacing. Cross-lingual code-switching replaces key trigger phrases with Swahili or English translations, fracturing the payload across linguistic boundaries and exploiting the monolingual training assumptions of conventional models, an especially potent strategy in the English-Swahili East African mobile money context.

\subsection{Attack Strategy --- Transformer Models}
The same attack strategy extends to the transformer tier, with one difference: rather than being applied generically, perturbations target only the tokens identified via attention weights as a heuristic estimate of token importance. This matters because preliminary experiments showed that generic black-box attacks on mBERT and XLM-RoBERTa produced negligible degradation (RDR $<$ 0.03)  at all intensities, confirming that untargeted perturbations underestimate the fragility of the transformer in the worst-case. The target tokens are identified by weighting the attention of the final encoder layer of the fine-tuned transformer, with the mean attention per token given by Equation ~\eqref{eq:attention}:

\begin{equation}
\bar{a}(t) = \frac{1}{H} \sum_{h=1}^{H} \text{Attention}_h(t)
\label{eq:attention}
\end{equation}
Tokens are ranked by Equation~\eqref{eq:attention}, and the same three perturbation strategies are applied exclusively to the top-$K$ ranked tokens, where $K = 3$ (low), $K = 6$ (medium), and $K = 10$ (high intensity).

\subsection{Adversarial Generation Quality Control}
Adversarial SMSs for model evaluation were generated using the Gemini API. Prompts followed a consistent structure comprising a neutral role framing positioning the model as a text encoder, precise task definition specifying the number of substitutions per intensity level, safeguards preserving phone numbers and names, two examples demonstrating correct output format, and a strict output constraint requiring the encoded string only, with no explanation or preamble. A manual readability check was performed on sampled messages per attack type to confirm that perturbations preserved human readability and produced natural-looking SMS content. Across attack types, perturbed messages retained plausible readability; code-switched messages in particular closely resembled genuine bilingual SMS traffic during manual inspection, confirming that this attack strategy blends naturally into the linguistic standards of the deployment context rather than introducing detectable artefacts.

\subsection{Evaluation Metric}
The primary evaluation metric is the Robustness Degradation Ratio (RDR), defined by Equation ~\eqref{eq:rdr}:
\begin{equation}
    \text{RDR} = \frac{F1_{\text{clean}} - F1_{\text{adversarial}}}{F1_{\text{clean}}}
    \label{eq:rdr}
\end{equation}
RDR is preferred over raw F1 because it measures relative rather than absolute degradation: two models can share an identical adversarial F1 but differ meaningfully with respect to robustness if their clean-text baselines differ, since the same F1 value represents a modest drop for a weak baseline but a severe collapse for a stronger one. Normalising against each model’s own clean-text performance isolates architectural fragility from baseline capability, supporting direct comparison between models. An RDR of 0 indicates absolute robustness; a value approaching 1 indicates catastrophic failure.

\section{Results}
\subsection{Clean-Text Baseline Performance}
Before adversarial evaluation, all models were evaluated in the unperturbed test set to establish clean-text baselines, reported in Table~\ref{tab:clean_baseline}.
\begin{table}
\centering
\caption{Clean-Text Baseline Performance}
\label{tab:clean_baseline}
\begin{tabular}{lc}
\hline
\textbf{Model} & \textbf{F1 Clean} \\
\hline
Random Forest  & 0.9410 \\
XGBoost        & 0.8950 \\
CNN+BiLSTM     & 0.8970 \\
mBERT          & 0.9429 \\
XLM-RoBERTa    & 0.9565 \\
\hline
\end{tabular}
\end{table}

XLM-RoBERTa achieved the highest clean-text F1 (0.9565), followed by mBERT (0.9429), Random Forest (0.941), CNN+BiLSTM (0.897), and XGBoost (0.895), establishing the baselines against which adversarial degradation is measured.

\subsection{Classical Model Results}
\begin{table}
\centering
\caption{Adversarial Results --- Classical Models}
\label{tab:classical_all}
\resizebox{\columnwidth}{!}{%
\begin{tabular}{llcccccc}
\hline
\multirow{2}{*}{\textbf{Model}} & \multirow{2}{*}{\textbf{Intensity}} & \multicolumn{2}{c}{\textbf{Char. Obfusc.}} & \multicolumn{2}{c}{\textbf{Structural}} & \multicolumn{2}{c}{\textbf{Code-Switch}} \\
 & & F1 Adv & RDR & F1 Adv & RDR & F1 Adv & RDR \\
\hline
\multirow{3}{*}{Random Forest}
  & Low    & 0.8892 & 0.0552 & 0.9245 & 0.0177 & 0.9212 & 0.0212 \\
  & Medium & 0.4333 & 0.5397 & 0.8224 & 0.1262 & 0.5366 & 0.4299 \\
  & High   & 0.0440 & 0.9533 & 0.0995 & 0.8943 & 0.2823 & 0.7001 \\
\hline
\multirow{3}{*}{XGBoost}
  & Low    & 0.8655 & 0.0330 & 0.8883 & 0.0076 & 0.8945 & 0.0007 \\
  & Medium & 0.4813 & 0.4623 & 0.4331 & 0.5162 & 0.6991 & 0.2189 \\
  & High   & 0.0110 & 0.9877 & 0.0766 & 0.9145 & 0.6726 & 0.2486 \\
\hline
\multirow{3}{*}{CNN+BiLSTM}
  & Low    & 0.8822 & 0.0160 & 0.8920 & 0.0051 & 0.8823 & 0.0170 \\
  & Medium & 0.8740 & 0.0252 & 0.7445 & 0.1696 & 0.8050 & 0.1021 \\
  & High   & 0.4601 & 0.4868 & 0.2496 & 0.7216 & 0.6561 & 0.2682 \\
\hline
\end{tabular}%
}
\end{table}

Table~\ref{tab:classical_all} shows that character obfuscation was the most destructive attack on the classical tier. At high intensity, Random Forest and XGBoost recorded RDR values of 0.9533 and 0.9877, reflecting near-total failure: homoglyph substitution corrupts the TF-IDF signal at the character level, leaving tokens unrecognisable to the learnt vocabulary. CNN+BiLSTM performed considerably better, reaching an RDR of 0.4868. The BiLSTM encoder retains some capacity to reconstruct meaning from tokens that escape substitution, which explains the degradation gap relative to bag-of-words models.

Structural perturbation produced a similar pattern, although the rankings shifted: XGBoost was most affected (RDR = 0.9145), marginally exceeding Random Forest (0.8943), as spacing manipulations shatter the $n$-gram boundaries upon which the trees are boosted. CNN+BiLSTM again showed resilience (0.7216), although degradation remained severe across all models, confirming that this attack is nearly as destructive as character obfuscation.

Code-switching produced the widest divergence within the tier. Random Forest recorded the highest degradation under code-switching, with an RDR of 0.7001. Its classification signal is highly dependent on the monolingual lexical frequency, so replacing trigger words precisely removes what the model relies on. XGBoost and CNN+BiLSTM degraded considerably less, with RDRs of 0.2486 and 0.2682, respectively: XGBoost draws on a broader feature distribution, while CNN+BiLSTM benefits from embedding-level proximity between substituted terms and their originals. All three models showed near-zero RDR at low intensities, with degradation rising sharply at medium and high intensities, thereby reproducing the non-linear threshold pattern observed across the classical tier.

\subsection{Transformer Model Results}
Table \ref{tab:transformer_all} presents the adversarial results for the transformer tier across the three attack types and intensity levels. Structural perturbation was the most damaging attack overall, surpassing character obfuscation at high intensity for both models, with RDR values of 0.3221 and 0.3511. Attention-guided targeting directs spacing manipulations toward the highest-weighted tokens, which disrupts the subword tokenisation both models rely on to process perturbed input. Character obfuscation produced a different pattern. Both models performed well at low and medium intensities, with RDR staying at or below 0.0541, but at high intensity, XLM-RoBERTa degraded more than mBERT, reaching 0.2326 compared to 0.1515, despite starting from a stronger clean-text baseline. That inversion supports the hypothesis that sharper attention concentration leaves decision-critical tokens more exposed under white-box access. It also reinforces the finding that clean-text performance does not reliably predict adversarial robustness.
\\

\begin{table}
\centering
\caption{Adversarial Results --- Transformer Models}
\label{tab:transformer_all}
\resizebox{\columnwidth}{!}{%
\begin{tabular}{llcccccc}
\hline
\multirow{2}{*}{\textbf{Model}} & \multirow{2}{*}{\textbf{Intensity}} & \multicolumn{2}{c}{\textbf{Char. Obfusc.}} & \multicolumn{2}{c}{\textbf{Structural}} & \multicolumn{2}{c}{\textbf{Code-Switch}} \\
 & & F1 Adv & RDR & F1 Adv & RDR & F1 Adv & RDR \\
\hline
\multirow{3}{*}{mBERT}
  & Low    & 0.8980 & 0.0476 & 0.8956 & 0.0502 & 0.9259 & 0.0180 \\
  & Medium & 0.8919 & 0.0541 & 0.7676 & 0.1859 & 0.9240 & 0.0201 \\
  & High   & 0.8000 & 0.1515 & 0.6392 & 0.3221 & 0.9178 & 0.0266 \\
\hline
\multirow{3}{*}{XLM-RoBERTa}
  & Low    & 0.9496 & 0.0072 & 0.9543 & 0.0023 & 0.9362 & 0.0156 \\
  & Medium & 0.9178 & 0.0405 & 0.8646 & 0.0961 & 0.9130 & 0.0400 \\
  & High   & 0.7340 & 0.2326 & 0.6206 & 0.3511 & 0.9051 & 0.0483 \\
\hline
\end{tabular}%
}
\end{table}

Code-switching produced the lowest degradation for both transformer models. The RDR peak reached 0.0483 for XLM-RoBERTa and 0.0266 for mBERT, both orders of magnitude below those of the other attack types. Multilingual pre-training positions equivalent cross-lingual expressions in proximate embedding regions, so translating a trigger phrase does not meaningfully displace its representation, even under attention-guided targeting. Transformer RDR stayed below 0.36 at its most severe. That ceiling remains well short of the near-total failure seen in classical models, and reflects the compounding effect of subword tokenisation and cross-lingual alignment, neither of which is unlimited on its own.

\begin{table}
\centering
\caption{Worst Case RDR --- All Models, All Attack Types}
\label{tab:peak_rdr_summary}
\resizebox{\columnwidth}{!}{%
\begin{tabular}{llccc}
\hline
\textbf{Tier} & \textbf{Model} & \textbf{Char. Obfusc.} & \textbf{Structural} & \textbf{Code-Switch} \\
\hline
\multirow{3}{*}{Classical}
  & Random Forest  & 0.9533 & 0.8943 & 0.7001 \\
  & XGBoost        & 0.9877 & 0.9145 & 0.2486 \\
  & CNN+BiLSTM     & 0.4868 & 0.7216 & 0.2682 \\
\hline
\multirow{2}{*}{Transformer}
  & mBERT          & 0.1515 & 0.3221 & 0.0266 \\
  & XLM-RoBERTa    & 0.2326 & 0.3511 & 0.0483 \\
\hline
\end{tabular}%
}
\end{table}

Table \ref{tab:peak_rdr_summary} summarises the worst case RDR across all models and attack types and provides strong evidence of an architectural boundary between the two tiers. Classical models show near-catastrophic degradation in the face of character obfuscation and structural perturbation, while transformer models remain substantially more resilient, with structural perturbation now being their worst-case attack rather than character obfuscation. Code-switching produces the largest intra-classical divergence, with Random Forest far exceeding CNN+BiLSTM, confirming that monolingual vocabulary bias is the decisive vulnerability factor for lexical architectures, while for transformers it remains the attack most fully neutralised by cross-lingual alignment. 

\begin{figure}[h]
\centering
\includegraphics[width=\columnwidth]{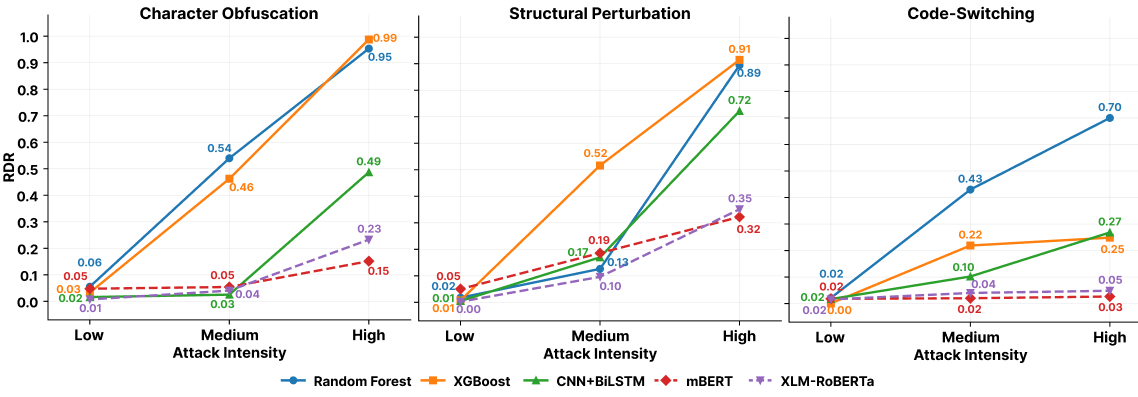}
\caption{RDR as a function of attack intensity across all five models and three attack types. Classical models (solid lines) escalate sharply past medium intensity; transformer models (dashed lines) remain comparatively flat except under structural perturbation.}
\label{fig:degradation_curves}
\end{figure}

Figure~\ref{fig:degradation_curves} makes this escalation visually explicit: classical models rise sharply past medium intensity, while transformers remain comparatively flat except under structural perturbation.

\subsection{Statistical Validation}
Two non-parametric tests confirm that observed RDR differences reflect genuine architectural properties rather than evaluation artefacts.

\subsubsection{Mann-Whitney U Test --- Cross-Tier Comparison}
A one-sided Mann-Whitney U test compared high-intensity RDR values between the classical and transformer architectures, treating each combination of model-attack as an independent observation ($n_1 = 9$ classical, $n_2 = 6$ transformer). The test statistic, given by Equation~\eqref{eq:mannwhitney}, counts the number of times a classical RDR value exceeds a transformer value in all possible pairings:
\begin{equation}
U = \sum_{i=1}^{n_1} \sum_{j=1}^{n_2} S(x_i, y_j), \quad
S(x_i, y_j) = \begin{cases} 1 & x_i > y_j \\
0.5 & x_i = y_j \\ 0 & x_i < y_j \end{cases}
\label{eq:mannwhitney}
\end{equation}
$U = 50.0$ ($p = 0.0024$) indicates that classical RDR exceeded transformer RDR in 50 of 54 pairings. The four exceptions occur where the worst-case attack of mBERT and XLM-RoBERTa (structural perturbation, RDR = 0.3221, 0.3511) exceeds the best-resisted attack of XGBoost and CNN+BiLSTM's (code-switching, RDR = 0.2486, 0.2682), a narrow ceiling-floor overlap rather than a separation breakdown. With $p < \alpha = 0.05$, the null hypothesis of Hypothesis~1 is rejected.

To quantify the magnitude of this separation, Cliff's delta was calculated directly from the same pairwise comparisons underlying $U$:
\begin{equation}
d = \frac{(\text{\# classical} > \text{transformer}) - (\text{\# classical} < \text{transformer})}{n_1 \times n_2}
\label{eq:cliffsdelta}
\end{equation}
Equation~\eqref{eq:cliffsdelta} gives $d = 0.85$, a large effect under conventional thresholds ($|d| \geq 0.474$), which confirms that separation is substantial rather than an artefact of sample size.

\subsubsection{Friedman Test --- Intensity Effect}
A Friedman test was applied per model using the three attack types as repeated observations of the intensity effect, with the statistic given by Equation~\eqref{eq:friedman}:
\begin{equation}
\chi^2_F = \frac{12}{nk(k+1)} \sum_{j=1}^{k} R_j^2 - 3n(k+1)
\label{eq:friedman}
\end{equation}

\noindent where $n = 3$ (attack types), $k = 3$ (intensity levels), and $R_j$ is the rank sum for the intensity level $j$. Random Forest, XGBoost, and CNN+BiLSTM each yielded $\chi^2 = 6.0$, $p = 0.050$, the consistent rank ordering (low $<$ medium $<$ high across all attacks) confirming the non-linear degradation threshold of Hypothesis~3 as an architectural property rather than an evaluation artefact.

The corresponding effect size, Kendall's coefficient of concordance, is:
\begin{equation}
W = \frac{\chi^2_F}{n(k-1)}
\label{eq:kendallsw}
\end{equation}

Equation~\eqref{eq:kendallsw} gives $W = 1.00$ for all three models, indicating a strong agreement in rank order between attack types. The marginal $p$-value reflects the small number of repeated measures ($n = 3$); the Friedman result is treated as supporting evidence alongside Table~\ref{tab:classical_all}.

\subsection{Addressing Research Questions}

The experimental results collectively address all three research questions.

\begin{itemize}
    \item \textbf{RQ1 (degradation under varying intensity):} Answered by the non-linear pattern observed across all classical models. RDR remains near-zero at low intensity but escalates sharply at medium and high intensity, with Random Forest and XGBoost reaching near-catastrophic failure at high intensity under both character obfuscation and structural perturbation. The Friedman test provides supporting evidence for this pattern as an architectural property rather than a sampling artefact ($\chi^2 = 6.0$, $p = 0.050$).

    \item \textbf{RQ2 (architectural determinants of differential robustness):} Answered by the clear tier separation. Transformer subword tokenisation and cross-lingual alignment provide compounding resilience advantages over TF-IDF surface matching, with classical models reaching near-catastrophic degradation at high intensity while transformers remain substantially more resilient across all attack types, a separation the Mann-Whitney U test confirms as statistically significant ($U = 50.0$, $p = 0.0024$, Cliff's $d = 0.85$).

    \item \textbf{RQ3 (whether clean-text performance predicts robustness):} Answered in negative. Performance-robustness decoupling is observed in both tiers, most notably where CNN+BiLSTM achieves the lowest clean-text F1 yet the lowest RDR within the classical tier, and XLM-RoBERTa, despite a higher clean-text baseline than mBERT, degrades more severely under high-intensity white-box targeting. Across both tiers, higher clean-text performance does not imply greater adversarial robustness.
\end{itemize}

\section{Architectural Failure Analysis}
\label{sec:failure_analysis}

The differential fragility observed across model classes is a direct consequence of the structural properties of each model's feature representation pipeline rather than incidental training variation.

TF-IDF models treat each word as an atomic token derived from its exact surface form; homoglyph substitution renders targeted tokens unrecognisable to the learnt vocabulary, collapsing their feature values to zero, and annihilating the discriminative signal entirely, explaining the near-catastrophic RDR values for Random Forest and XGBoost under character obfuscation. CNN+BiLSTM's lower RDR reflects partial contextual recovery via its BiLSTM encoder, though this recovery is overwhelmed at high intensity. Transformer models resist this attack through subword tokenisation, decomposing obfuscated surface forms into recognisable subword units and preserving partial semantic signal; the robustness inversion at high intensity, where XLM-RoBERTa degrades more severely than mBERT despite its superior baseline, is consistent with the hypothesis that XLM-RoBERTa's sharper attention concentration makes its decision-critical tokens more exploitable under white-box targeting.

A similar mechanism underlies the vulnerability to structural perturbation: the statistical identity of TF-IDF $n$-gram features depends on the exact alignment of the character boundary, and spacing manipulations shatter these boundaries, mapping targeted tokens to low-weight fragments absent from the learnt feature space. XGBoost is most sensitive to this disruption, crossing the high-risk threshold at medium intensity, while CNN+BiLSTM's convolutional filters, operating over embedding sequences rather than discrete $n$-gram counts, provide partial insulation, although degradation remains severe at high intensity.

Code-switching exploits a third distinct mechanism: TF-IDF models assign discriminative weight to language-specific surface forms independently, treating semantically equivalent cross-lingual expressions as entirely unrelated features. Code-switching removes high-weight trigger tokens from the input vector without replacement, collapsing detection confidence, most severely for Random Forest, reflecting the highest degree of monolingual lexical concentration. Transformer models prove nearly immune to this attack, as multilingual pre-training maps semantically equivalent cross-lingual expressions into proximate embedding regions; code-switching alters surface realisation without removing the discriminative signal, explaining the largest cross-architectural divergence observed in this study.

Table~\ref{tab:peak_rdr_summary} makes the boundary explicit: surface-form lexical matching produces near-catastrophic failure under character obfuscation and structural perturbation, while CNN+BiLSTM's lower RDR confirms contextual embeddings offer partial but bounded insulation, and the robustness inversion reaffirms that clean-text performance does not predict adversarial robustness.

\section{Defence Considerations}
\label{sec:defence_consideration}
The failure modes identified in Section~\ref{sec:failure_analysis} suggest different architecture-specific mitigations. These countermeasures are proposed
theoretically; empirical validation is reserved for future work.

\subsection{Classical Model Defences}
For TF-IDF-based models, \textit{Unicode normalisation preprocessing} would canonicalise homoglyphs before feature extraction, countering character obfuscation. \textit{Character n-gram augmentation} reduces sensitivity to spacing perturbations, and \textit{bilingual vocabulary extension} addresses the monolingual bias exploited by code-switching attacks. For CNN+BiLSTM, \textit{adversarial training augmentation} is the most practical mitigation, leveraging its existing sequential context encoder without architectural changes.

\subsection{Transformer Model Defences}
\textit{Attention-aware adversarial training} is hypothesised to distribute attention weight more broadly, reducing the concentration exploited under white-box targeting of XLM-RoBERTa. \textit{Input-layer Unicode sanitisation} before subword tokenisation further reduces the attack surface for character obfuscation across both transformer models.

\section{Limitations}
Several constraints are inscribed in these findings. The malicious-class test set is small (356 messages overall, approximately 71 held out), a consequence of the low-resource setting under investigation; the Friedman test's consistent rank ordering mitigates but does not eliminate this concern, and larger annotated corpora remain necessary for further validation. The study is limited to English-Swahili; although the attack strategies are designed to be language-agnostic, empirical validation has not been performed in other pairs of low-resource languages. Adversarial examples were generated via the Gemini API rather than collected from real-world campaigns, and synthetic generation may not fully capture attackers' real-world tactics despite passing readability verifications. Finally, attention-based targeting assumes that attention weight reflects the tokens most influential to a model's decision; this is a widely used but imperfect proxy, since attention does not necessarily correspond to causal influence, and validation against alternative targeting strategies is left to future work. The results should therefore be interpreted as evidence of comparative robustness patterns rather than precise estimates of deployment-time attack success.

\section{Future Work}
Three directions follow from these findings: evaluating adversarial training augmentation as a cost-effective robustness route for embedding-based classical architectures (e.g., CNN+BiLSTM) where transformer fine-tuning is prohibitive; developing and testing attention-aware adversarial training for XLM-RoBERTa to assess whether it reduces vulnerability to targeted character obfuscation; and constructing combined multi-tactic threat models reflecting adversaries that mix strategies (e.g., character obfuscation on high-attention tokens coupled with code-switching).

Three methodological extensions are also planned: validating attention-based targeting against a random-token baseline to confirm that attention weight identifies genuinely more exploitable tokens; empirically evaluating at least one proposed defence strategy (Section~\ref{sec:defence_consideration}) rather than treating them as purely theoretical; and reporting bootstrap confidence intervals around RDR estimates to quantify uncertainty given the limited malicious test set.

\section{Conclusion}
This research highlights a distinct architectural divide in adversarial smishing detection. Traditional models experience drastic performance drops under character obfuscation and structural alterations (RDR $>$ 0.89). In contrast, transformer architectures exhibit greater resilience (RDR $\leq$ 0.35, primarily struggling with structural perturbations as their worst-case vulnerability. In particular, a robustness inversion occurs across attacks: Despite a higher clean-text accuracy, XLM-RoBERTa proves to be less resilient than mBERT, demonstrating that baseline performance does not guarantee adversarial defence.

The uncovered vulnerabilities—including n-gram reliance, tokenisation granularity, and monolingual bias—offer a framework for designing secure systems and prompt further investigation into multi-tactic attacks and adversarial training. In resource-constrained environments where transformers are computationally impractical, the CNN+BiLSTM architecture serves as the most viable classical alternative. Ultimately, security professionals must assess models based on adversarial resilience rather than baseline metrics, as models with comparable clean-text performance often exhibit starkly different real-world defensive capabilities.

\bibliographystyle{splncs04}
\bibliography{315-ISBM-2026}

@inproceedings{conneau_unsupervised_2020,
	title = {Unsupervised {Cross}-lingual {Representation} {Learning} at {Scale}},
	language = {en},
	urldate = {2026-07-29},
	author = {Conneau, Alexis and Khandelwal, Kartikay and Goyal, Naman and Chaudhary, Vishrav and Wenzek, Guillaume and Guzmán, Francisco and Grave, Edouard and Ott, Myle and Zettlemoyer, Luke and Stoyanov, Veselin},
	year = {2020},
	pages = {8440--8451},
}

@misc{boucher_bad_2021,
	title = {Bad {Characters}: {Imperceptible} {NLP} {Attacks}},
	shorttitle = {Bad {Characters}},
	doi = {10.48550/arXiv.2106.09898},
	language = {en},
	urldate = {2026-07-29},
	publisher = {arXiv},
	author = {Boucher, Nicholas and Shumailov, Ilia and Anderson, Ross and Papernot, Nicolas},
	month = dec,
	year = {2021},
}

@article{xu_malicious_2025,
	title = {Malicious {SMS} detection using ensemble learning and {SMOTE} to improve mobile cybersecurity},
	volume = {154},
	issn = {01674048},
	language = {en},
	urldate = {2026-03-04},
	journal = {Computers \& Security},
	author = {Xu, Hongsheng and Qadir, Akeel and Sadiq, Saima},
	month = jul,
	year = {2025},
	pages = {104443},
}

@inproceedings{tan_code-mixing_nodate,
	title = {Code-{Mixing} on {Sesame} {Street}: {Dawn} of the {Adversarial} {Polyglots}},
	booktitle = {Proceedings of the 2021 {Conference} of the {North} {American} {Chapter} of the {Association} for {Computational} {Linguistics}: {Human} {Language} {Technologies}},
	publisher = {Association for Computational Linguistics},
	author = {Tan, Samson and Joty, Shafiq},
	pages = {3596--3616},
}

@article{shaikh_smishing_2025,
	title = {Smishing {Detection}: {Combating} {SMS} {Phishing} {Attacks} by {Utilizing} {Machine}-{Learning} {Algorithms}},
	volume = {14},
	issn = {22783075},
	shorttitle = {Smishing {Detection}},
	language = {en},
	number = {5},
	urldate = {2026-05-01},
	journal = {International Journal of Innovative Technology and Exploring Engineering},
	author = {Shaikh, Aqsa and Shaikh, Mariya and {Srivaramangai R.}},
	month = apr,
	year = {2025},
	pages = {28--33},
}

@article{sankaine_english-swahili_2025,
	title = {An {English}-{Swahili} {Email} {Spam} {Detection} {Model} for {Improved} {Accuracy} {Using} {Convolutional} {Neural} {Networks}},
	volume = {5},
	copyright = {Copyright (c) 2025 Leshan  Sankaine, John G.  Ndia , Dennis  Kaburu},
	issn = {2958-6542},
	language = {en},
	number = {2},
	urldate = {2026-02-19},
	journal = {Mesopotamian Journal of CyberSecurity},
	author = {Sankaine, Leshan and Ndia, John G. and Kaburu, Dennis},
	month = jun,
	year = {2025},
	pages = {590--605},
}

@inproceedings{pires_how_2019,
	address = {Florence, Italy},
	title = {How {Multilingual} is {Multilingual} {BERT}?},
	language = {en},
	urldate = {2026-03-11},
	booktitle = {Proceedings of the 57th {Annual} {Meeting} of the {Association} for {Computational} {Linguistics}},
	publisher = {Association for Computational Linguistics},
	author = {Pires, Telmo and Schlinger, Eva and Garrette, Dan},
	year = {2019},
	pages = {4996--5001},
}

@misc{morris_textattack_2020,
	title = {{TextAttack}: {A} {Framework} for {Adversarial} {Attacks}, {Data} {Augmentation}, and {Adversarial} {Training} in {NLP}},
	shorttitle = {{TextAttack}},
	language = {en},
	urldate = {2026-03-04},
	publisher = {arXiv},
	author = {Morris, John X. and Lifland, Eli and Yoo, Jin Yong and Grigsby, Jake and Jin, Di and Qi, Yanjun},
	month = oct,
	year = {2020},
	note = {arXiv:2005.05909 [cs]},
}

@article{mambina_uncovering_2024,
	title = {Uncovering {SMS} {Spam} in {Swahili} {Text} {Using} {Deep} {Learning} {Approaches}},
	volume = {12},
	issn = {2169-3536},
	urldate = {2025-11-13},
	journal = {IEEE Access},
	author = {Mambina, Iddi S. and Ndibwile, Jema D. and Uwimpuhwe, Deo and Michael, Kisangiri F.},
	year = {2024},
	pages = {25164--25175},
}

@article{mambina_classifying_2022,
	title = {Classifying {Swahili} {Smishing} {Attacks} for {Mobile} {Money} {Users}: {A} {Machine}-{Learning} {Approach}},
	volume = {10},
	issn = {2169-3536},
	shorttitle = {Classifying {Swahili} {Smishing} {Attacks} for {Mobile} {Money} {Users}},
	urldate = {2025-11-13},
	journal = {IEEE Access},
	author = {Mambina, Iddi S. and Ndibwile, Jema D. and Michael, Kisangiri F.},
	year = {2022},
	pages = {83061--83074},
}

@inproceedings{li_textbugger_2019,
	address = {San Diego, CA},
	title = {{TextBugger}: {Generating} {Adversarial} {Text} {Against} {Real}-world {Applications}},
	isbn = {978-1-891562-55-6},
	shorttitle = {{TextBugger}},
	language = {en},
	urldate = {2026-03-04},
	booktitle = {Proceedings 2019 {Network} and {Distributed} {System} {Security} {Symposium}},
	publisher = {Internet Society},
	author = {Li, Jinfeng and Ji, Shouling and Du, Tianyu and Li, Bo and Wang, Ting},
	year = {2019},
}

@article{hosseinpour_complex-network_2024,
	title = {Complex-network based model for {SMS} spam filtering},
	volume = {255},
	issn = {13891286},
	language = {en},
	urldate = {2026-03-04},
	journal = {Computer Networks},
	author = {Hosseinpour, Shaghayegh and Shakibian, Hadi},
	month = dec,
	year = {2024},
	pages = {110892},
}

@article{haizam_analysing_2024,
	title = {Analysing {The} {Impact} of {Smishing} {Attack} in {Public} {Announcement} {System} on {Mobile} {Phone}},
	volume = {245},
	issn = {18770509},
	language = {en},
	urldate = {2026-03-04},
	journal = {Procedia Computer Science},
	author = {Haizam, Mohamad Nurhafiz Bin and Zulkipli, Nurul Huda Binti Nik},
	year = {2024},
	pages = {1165--1174},
}

@inproceedings{bajaj_deceiving_2023,
	title = {Deceiving {Deep} {Learning}-based {Fraud} {SMS} {Detection} {Models} through {Adversarial} {Attacks}},
	urldate = {2026-03-04},
	booktitle = {2023 17th {International} {Conference} on {Signal}-{Image} {Technology} \& {Internet}-{Based} {Systems} ({SITIS})},
	author = {Bajaj, Ashish and Vishwakarma, Dinesh Kumar},
	month = nov,
	year = {2023},
	pages = {327--332},
}

@article{asheshemi_adversarial_2025,
	title = {Adversarial {Robustness} in {Natural} {Language} {Processing}: {An} {Empirical} {Analysis} of {Machine} {Learning} {Model} {Vulnerabilities} to {Adversarial} {Attacks}},
	volume = {10},
	journal = {International Journal of Research and Innovation in Applied Science},
	author = {Asheshemi, Nelson and Daniel, Okoro and Micheal, Obode},
	month = nov,
	year = {2025},
	pages = {1921--1939},
}

\end{document}